\documentclass[preprint,review,12pt]{elsarticle}
\usepackage{graphicx}
\usepackage{subfigure}

\usepackage{amssymb}
\usepackage{amsthm}

 \biboptions{sort&compress}

\journal{Opt. Commun.}

\begin{document}

\begin{frontmatter}



\title{Transmission and Reflection through $1$D Metallo-Dielectric Gratings of Real Metals under Sub-wavelength Condition}


\author[1,*]{A. T. M. Anishur Rahman}\author[1]{Peter Majewski}\author[2]{ Krasimir Vasilev}  

\address[1]{School of AME, University of South Australia, Mawson Lakes, SA 5095, Australia} \address[2]{Mawson Institute and School of AME, University of South Australia, Mawson Lakes, SA 5095, Australia}
\address[*]{Corresponding author: rahaa001@mymail.unisa.edu.au}

\begin{abstract}
Under the sub-wavelength condition ($w<\lambda/2$), an analytical model of light transmission and reflection through $1$D metallo-dielectric gratings of real metals has been developed. It has been shown that the transmission intensity associated with the Fabry-Perot (FP) resonance of a $1$D metallo-dielectric grating of a real metal decreases with the increasing grating thickness and the dielectric constant of the ridge material. Further, it has also been demonstrated that the intensity of the FP resonance increases with the increasing slit width while it is independent of the grating period ($P$) and the incidence angle (when $P << \lambda$).

\end{abstract}


\end{frontmatter}


\section{Introduction}

\begin{figure}[h]
\centering
\includegraphics[width=7.5cm]{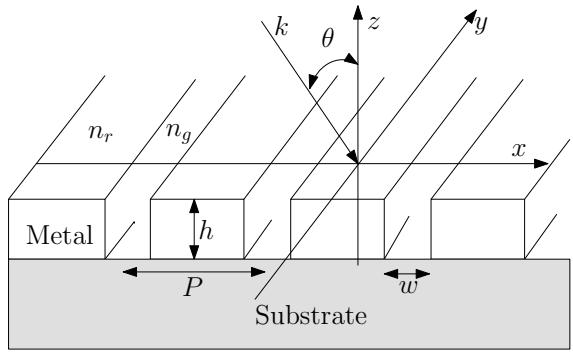}
\centering
\caption{$1$D Lamellar Grating}
\label{fig1_chap4}
\end{figure}

After the discovery of Extraordinary Optical Transmission (EOT) through $1$D metallo-dielectric gratings, a number of efforts, both numerical and semi-analytical, has been undertaken to understand the physics behind this exceptional phenomenon \cite{Porto1999,Astilean2000,Lalanne2000A,Popov2000,GarciaVidal2002,Cao2002,Steele2003,Marquier2005,Garcia2007}. Numerical approaches of modeling EOT devices include the transfer matrix formalism \cite{Pendry1994,Bell1995,Porto1999,GarciaVidal2002}, the Finite Difference Time Domain (FDTD) method \cite{Taflove2005,Garcia2007}, the Rigorous Coupled Wave Analysis (RCWA) \cite{Moharam1995,Lalanne2000A,Cao2002,Marquier2005}, and the Exact Modal Analysis (EMA) \cite{Sheng:82}. Both the transfer matrix and FDTD methods solve Maxwell's equations directly using grids and produce accurate results. Other features of these two methods include the ability of visualizing electric and magnetic fields in a plane, both inside and outside of a structure understudy. Further, flow of energy (Poynting vector) at any location at a particular instant of time can also be calculated and monitored. However, the transfer matrix formalism and the FDTD method are computationally very demanding \cite{Garcia2010}. In contrast, the RCWA and the EMA are the two closely related numerical techniques which are widely used to analyze $1$D structures \cite{Astilean2000,Lalanne2000A,Collin2001,Cao2002, GarciaVidal2002,Marquier2005}. The RCWA expresses the dielectric function of the overall grating structure and the Electromagnetic (EM) field inside the grating region ($-h\le z \le 0$, see Fig. \ref{fig1_chap4}) in terms of Fourier series \cite{Moharam1995}. With the help of Maxwell's equations a space harmonic of the EM field can be expressed as a function of the remaining harmonics of the EM field and harmonics of the dielectric function. After applying proper boundary conditions at $z=0$ and $z=-h$ interfaces, reflection and transmission efficiencies of various diffraction orders can be calculated \cite{Moharam1995}. In the case of EMA, electromagnetic fields inside the grating region are expressed as waveguide modes. Profiles and wave numbers of various modes are determined numerically by solving a transcendental equation obtained using the stratified medium theory \cite{Sheng:82}. Once profiles and wave numbers of different modes are available, Maxwell's equations in combination with boundary conditions (those like the RCWA) are solved to find reflection and transmission coefficients. At this point it is relevant to mention that although the numerical techniques discussed here produce good results, they are devoid of physical insights. Specifically, relationships among transmission and reflection coefficients, and $w$, $P$ and $h$ are not clearly understood. Further, simulation techniques discussed above require significant computational power.

To avoid numerical simulation, different simplified models of light transmission and reflection through $1$D gratings under sub-wavelength ($w<0.5\lambda/n_g$, where $n_g$ is the dielectric constant of the slits) condition have been developed \cite{Porto1999,GarciaVidal2002,Lalanne2000A}. In developing these models, the fundamental waveguide mode is the only propagating mode inside the slit/cavity is assumed. In addition, these models also assume that the grating ridge and/or the slit walls are made of Perfect Electrical Conductor (PEC) \cite{Porto1999,GarciaVidal2002,Lalanne2000A}. For most of the metals the PEC assumption holds true in the deep infrared and microwave regions of the electromagnetic spectrum and that is where the previous models produce reliable results. In the visible band of the electromagnetic spectrum, metals become significantly lossy \cite{Marquier2005,Billaudeau2009} and the PEC assumption along with the existing transmission and reflection models of light fail. In this context, a model which does not assume the PEC assumption is necessary to analyze $1$D grating structure in the visible band of the electromagnetic spectrum. Further, a model of $1$D metallo-dielectric gratings which can provide explanation of light transmission and reflection in terms of grating geometrical parameters is also relevant considering the limitations of the existing models. In this article, we aim to develop a model of $1$D metallo-dielectric gratings of real metals which complements the shortcomings of the current semi-analytical (involves infinite sums) models. In developing this model, $w<\lambda/(2n_g)$ is assumed.

\section{One mode model of $1$D transmission grating}
Let us assume a TM polarized electromagnetic wave $H_{inc,y}=exp{\{-i\omega t\}}*\exp{\{in_1k_0(\sin\theta x-\cos\theta z)\}}$ of wavelength $\lambda$ is incident upon the grating structure shown in Fig. \ref{fig1_chap4} at an incident angle $\theta$, where $n_1$ is the refractive index of the incident medium ($z>0$) and $k_0=2\pi/\lambda$ is the free space wave number. $n_g$ and $n_r$ are the refractive indexes of the grooves/slits and the grating ridges respectively. The dielectric constants of the slits and ridges can be written as $\epsilon_g=n_g^2$ and $\epsilon_r=n_r^2$ correspondingly. A part of the incident energy is reflected back to the incident medium as diffraction orders whereas another part emerges as transmitted diffraction orders in the substrate ($-h<z$) of refractive index $n_s$. Using Rayleigh expansion the reflected and transmitted diffraction orders can be expressed as -

\begin{eqnarray}
\nonumber
H_y^r&=&H_{inc,y}+\sum_mR_m\exp{[ik_0\{\gamma_m x+(n_1^2-\gamma_m^2)^{1/2}z\}]}\\
H_y^t&=&\sum_mT_m\exp{[ik_0\{\gamma_m x-(n_s^2-\gamma_m^2)^{1/2}z\}]}
\label{eq1_chap4}
\end{eqnarray}

where $\gamma_m=n_1\sin{\theta}+m\lambda/P$ and $m$ is an integer. $R_m$ and $T_m$ are the reflected and transmitted magnetic field amplitudes \cite{Sheng:82}.

The transmitted energy passes through the slits from the incident medium to the substrate as waveguide modes. Based upon $\lambda$, $w$, and $P$, different waveguide modes are excited inside the slits. Mode profiles and the $x-$ and $z-$ components of wave vectors of different waveguide modes can be found by solving Maxwell's equations in the grating region ($-h <z<0$) in combination with the stratified medium theory \cite{Sheng:82}. The $x-$ and $z-$ components of various waveguide modes can be found from Eq. (\ref{eq2_chap4}).

\begin{eqnarray}
\small
\cos(k_0n_1P\sin\theta)-\cos(\beta tP)\cos(\alpha
fP)+
\frac{1}{2}[\frac{\epsilon_r\alpha}{\beta}+
\frac{\beta}{\epsilon_r\alpha}]\sin(\beta
tP)\sin(\alpha fP)=0
\label{eq2_chap4}
\end{eqnarray}

where $t=(P-w)/P$, $f=1-t=w/P$, and $\alpha=k_0\sqrt{\epsilon_g-n_{Eff}^2}$ and $\beta=k_0\sqrt{\epsilon_r-n_{Eff}^2}$ are the $x-$ components of a waveguide mode in the slit/groove and ridge materials respectively. The $z-$ component of a waveguide mode is defined as $k_z=n_{Eff}k_0$. For metallic ridges, the dielectric constant is given by $\epsilon_r=n_r^2=(\eta+i\kappa)^2$, where $\eta$ and $\kappa$ are the real and imaginary components of the refractive index, respectively. As can be understood from Eq. (\ref{eq2_chap4}) that there are infinite number of solutions of this equation and each of these solutions is called a waveguide mode. However, when $w<\lambda/2$ with $n_g=1$, only the fundamental waveguide mode is propagating inside a slit cavity \cite{Porto1999,Astilean2000,Lalanne2000A}. In this case, given that $|n_r^2|>>|n_{Eff}^2|$, $n_{Eff}$ of the fundamental mode can be expressed as Eq. (\ref{eq3_chap4}) \cite{Rahman2011A}.

\begin{equation}
n_{Eff}^2=1+i\frac{\lambda}{\pi w n_r}
\label{eq3_chap4}
\end{equation}

Considering Eq. (\ref{eq3_chap4}), $\alpha$ and $\beta$ can be expressed as $k_0\sqrt{-i\lambda/(\pi w n_r)}$ and $k_0n_r$ respectively. Mode profile of the fundamental waveguide mode can be found from Eq. (\ref{eq4_chap4}) \cite{Sheng:82}.

\begin{eqnarray}
 \nonumber
 X(x)&=&U_1\cos{[\alpha(x-\frac{t P}{2})]}+j\frac{V_1\epsilon_g k_0}{\alpha}\sin{[\alpha(x-\frac{t P}{2})]},\hspace{0.25cm}\frac{t P}{2}\le |x|\le (1-\frac{t}{2})P\\
 &=&\cos{[\beta(x+\frac{t P}{2})]}+j\frac{V_0\epsilon_r k_0}{\beta}\sin{[\beta(x+\frac{t P}{2})]},\hspace{0.75cm}|x|\le \frac{t P}{2}
 \label{eq4_chap4}
 \end{eqnarray}

 where $V_0$, $V_1$, $U_1$, $M$ and $N$ are give by

 \begin{eqnarray}
 V_0&=&[exp(jk_0P\sin{\theta})-M]/N\\
 V_1&=&j\frac{\beta}{k_0\epsilon_r}\sin{(\beta t P)}+V_0\cos{(\beta t P)}\\
 U_1&=&\cos{(\beta t P)}+j\frac{k_0V_0\epsilon_r}{\beta}\sin{(\beta t P)}\\
 M&=&\cos{(\beta t P)}\cos{(\alpha f P)}-\frac{\beta}{\epsilon_r\alpha}\sin{(\beta t P)}\sin{(\alpha f P)}\\
 N&=&jk_0[\cos{(\beta t P)}\sin{(\alpha f P)}/\alpha+\epsilon_r\sin{(\beta t P)}\cos{(\alpha f P)}/\beta]
 \end{eqnarray}

Given that $|\sqrt{-i4\pi w/(n_r\lambda)}| << 1$, $|\cos{\sqrt{-i4\pi w/(n_r\lambda)}}|\approx 1$, $\hspace{1cm}$ and $|\sin{\sqrt{-i4\pi w/(n_r\lambda)}}|\approx 0$, $M$ and $N$ can be expressed as below -.

\begin{eqnarray}
M&\approx&\cos{\{k_0(P-w)n_r\}}\cos{(\sqrt{-i\frac{4\pi w}{\lambda n_r}})}\\
N&\approx &i n_r\sin{\{k_0(P-w)n_r\}}\cos{(\sqrt{-i\frac{4\pi w}{\lambda n_r}})}
\end{eqnarray}

Further, provided that $|\exp{\{ik_0P\sin{\theta}\}}| = 1$ and $|N|>>1$, $V_0$, $V_1$ and $U_1$ can be expressed as below -.

\begin{eqnarray}
V_1&\approx &0\\
U_1 &\approx& 0\\
V_0&\approx & i\frac{\cot{\{k_0(P-w)n_r\}}}{n_r}
\end{eqnarray}

Substituting $M$, $N$, $V_0$, $V_1$, $U_1$, $\alpha$ and $\beta$ in Eq. (\ref{eq4_chap4}), $X(x)$ corresponding to the fundamental mode can be expressed as below -.

\begin{eqnarray}
\nonumber
X(x) &\approx &\cos{\{k_0(x+\frac{P-w}{2})n_r\}}-\cot{\{k_0(P-w)n_r\}}\sin{\{k_0(x+\frac{P-w}{2})n_r\}}\\\nonumber &&\hspace{5cm} -\frac{tP}{2} \le x\le \frac{tP}{2}\\
X(x) &\approx & 0 \hspace{5cm}\frac{tP}{2}\le |x|\le (1-\frac{t}{2})P
\label{eq5_chap4}
\end{eqnarray}

Provided that the fundamental waveguide mode is the only surviving mode, the magnetic field inside the grating region ($-h<z<0$) can be expressed as

\begin{equation}
H_{gy}=X(x)[A\exp{(in_{Eff}k_0z)}+B\exp{(-in_{Eff}k_0z)}]
\label{eq6_chap4}
\end{equation}

where $A$ and $B$ are the co-efficients of the counter propagating fundamental mode inside the slit cavity. After applying boundary conditions at $z=0$ and $z=-h$ interfaces for the magnetic and electric fields in combination with Maxwell's equations, one gets relations involving $A$, $B$, $R_m$ and $T_m$ as follows -.
\begin{eqnarray}
(A+B)X_m-R_m=\delta_{m0}\\
\label{eq7_chap4}
(A-B)\Omega_m-\Pi_m R_m=-\Pi_m\delta_{m0}\\
\label{eq8_chap4}
(A\Sigma+B\Sigma^{-1})X_m-\Delta_m T_m=0\\
\label{eq9_chap4}
(A\Sigma-B\Sigma^{-1})\Omega_m+\Delta_m\xi_m T_m=0
\label{eq10_chap4}
\end{eqnarray}

where $\Sigma=exp(-ihk_z)$ and $\Sigma^{-1}=exp(ihk_z)$ are constants. $k_z=n_{Eff}k_0$ is the $z$-component of the wave vector corresponding to the fundamental waveguide mode. $\Pi_m=k_0\sqrt{(n_1^2-\gamma_{m}^2)}$, $\xi_m=k_0\sqrt{(n_s^2-\gamma_{m}^2)}/n_s$, and $\Delta_m=\exp{(ik_0h\sqrt{n_s^2-\gamma_{m}^2})}$. $X_m=(1/P)\int_{-Pt/2}^{(1-t/2)P}X(x) \exp{(-ik_0\gamma_{m} x)}dx$, $\Omega_m = (k_z/P)\int_{-Pt/2}^{(1-t/2)P}\Omega(x)\exp{(-ik_0\gamma_{m} x)}dx$, and $\Omega(x)=X(x)/\epsilon$. For $-tP\le x\le tP$, $\epsilon$ is equal to $\epsilon_r$ whereas for $tP\le x \le (1-\frac{t}{2})P$, $\epsilon$ is $n_g^2=1$.

After some simple algebraic manipulation of equations (\ref{eq7_chap4})-(\ref{eq10_chap4}) and assuming $n_1=1$, $T_m$ and $R_m$ can be written as follows.

\begin{eqnarray}
\small
T_m=\frac{2\delta_{m0}\exp{(-ik_0h\sqrt{n_s^2-\gamma_m^2})}}
{[1+\frac{1}{n_s}\sqrt{\frac{n_s^2-\gamma_m^2}{1-\gamma_m^2}}]\cos{(n_{Eff}k_0h)}-i[\frac{\epsilon_r\sqrt{n_s^2-\gamma_m^2}}{n_sn_{Eff}}+\frac{n_{Eff}}{\epsilon_r\sqrt{1-\gamma_m^2}}]\sin{(n_{Eff}k_0h)}}
\label{eq13_chap4}
\end{eqnarray}

\begin{eqnarray}
\nonumber
R_m=\frac{2\delta_{m0}[\cos{(n_{Eff}k_0h)}-i\frac{\epsilon_r}{n_{Eff}}\sqrt{1-\frac{\gamma_m^2}{n_s^2}}\sin{(n_{Eff}k_0h)}] }{[1+\frac{1}{n_s}\sqrt{\frac{n_s^2-\gamma_m^2}{1-\gamma_m^2}}]\cos{(n_{Eff}k_0h)}-i[\frac{\epsilon_r\sqrt{n_s^2-\gamma_m^2}}{n_sn_{Eff}}+\frac{n_{Eff}}{\epsilon_r\sqrt{1-\gamma_m^2}}]\sin{(n_{Eff}k_0h)}}\\-\delta_{m0}\hspace{3cm}
\label{eq14_chap4}
\end{eqnarray}

 By expressing $n_{Eff}$ as $n_{Eff}^{'}+i n_{Eff}^{''}$, and given that $\sinh{(n_{Eff}^{''}k_0h)} <<1$ (since $n_{Eff}^{''} << 1$) and $|\frac{n_{Eff}}{\epsilon_r}|<< 1$, for $m=0$, $T_0$ and $R_0$ can be simplified to -

\begin{eqnarray}
\nonumber
T_0=\frac{2\exp{(-ik_0h\sqrt{n_s^2-\sin^2{\theta}})}}
{[(1+\frac{1}{n_s}\sqrt{\frac{n_s^2-\sin^2{\theta}}{1-\sin^2{\theta}}})\cos{(n_{Eff}^{'}k_0h)}-i\frac{\epsilon_r\sqrt{n_s^2-\sin^2{\theta}}}{n_sn_{Eff}}\sin{(n_{Eff}^{'}k_0h)}]}\\\times{\frac{1}{\cosh{(n_{Eff}^{''}k_0h)}}}\hspace{3cm}
\label{eq15_chap4}
\end{eqnarray}

\begin{eqnarray}
R_0=\frac{2[\cos{(n_{Eff}^{'}k_0h)}-i\frac{\epsilon_r\sqrt{n_s^2-\sin^2{\theta}}}{n_sn_{Eff}}\sin{(n_{Eff}^{'}k_0h)}]}
{(1+\frac{1}{n_s}\sqrt{\frac{n_s^2-\sin^2{\theta}}{1-\sin^2{\theta}}})\cos{(n_{Eff}^{'}k_0h)}-i\frac{\epsilon_r\sqrt{n_s^2-\sin^2{\theta}}}{n_sn_{Eff}}\sin{(n_{Eff}^{'}k_0h)}}-1
\label{eq16_chap4}
\end{eqnarray}

It can be seen from Eqs. (\ref{eq13_chap4}) and (\ref{eq14_chap4}) that unlike the existing semi-analytical models \cite{Porto1999,GarciaVidal2002,Lalanne2000A} of light transmission and reflection through the $1$D grating structure, the models presented here are analytical and hence the physics of light transmission and reflection can be easily understood. In the next section different examples are considered and other benefits of the proposed models are discussed.

\section{Results and Discussion}

\begin{figure}
\centering
\subfigure{
\includegraphics[width=7.5cm]{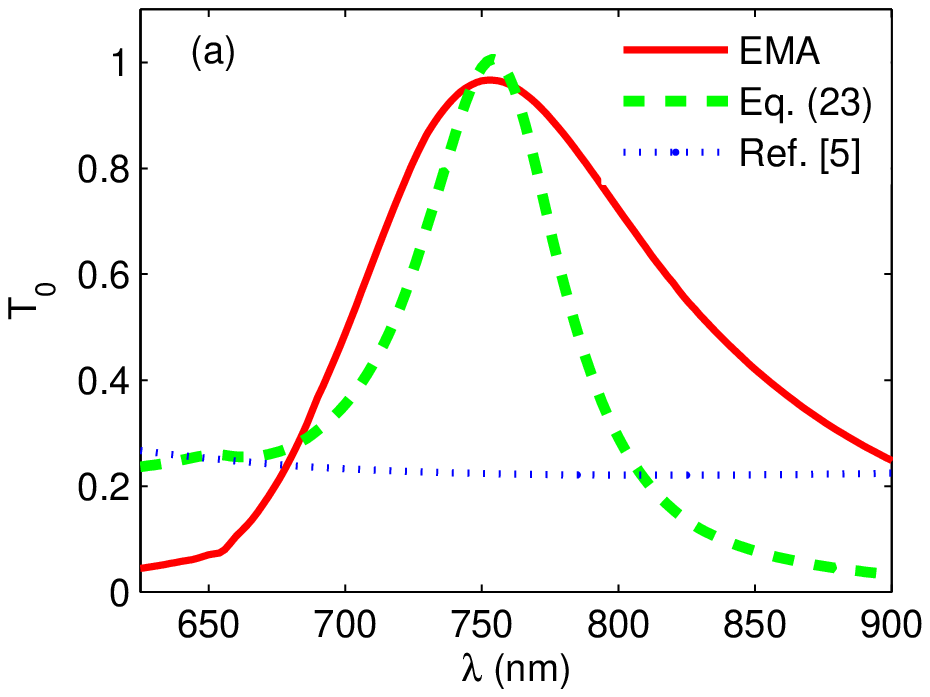}}
\subfigure{
\includegraphics[width=7.5cm]{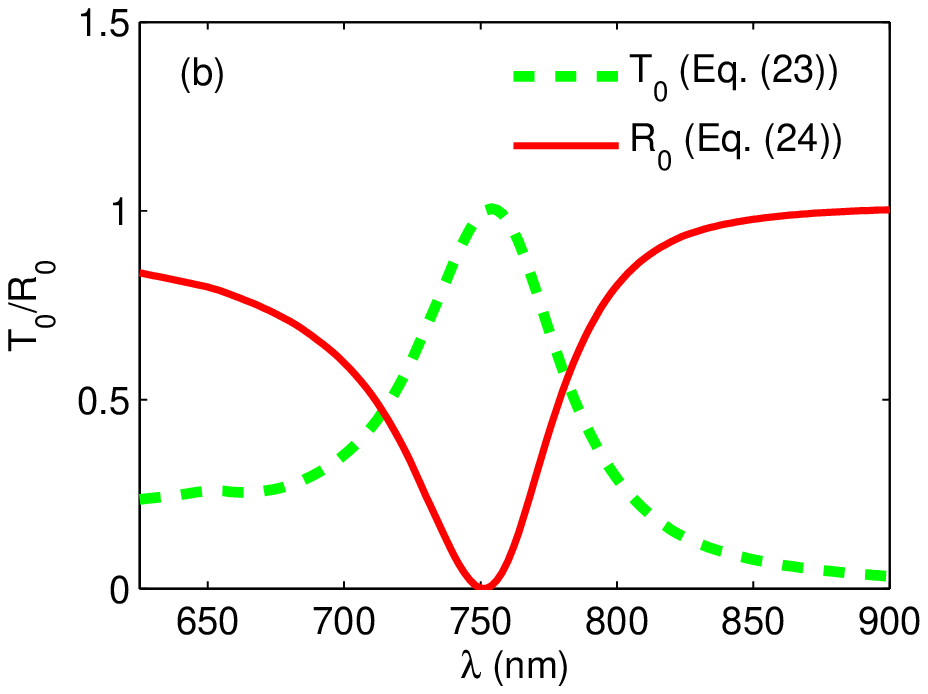}}
\centering
\caption{Zeroth order (a) transmission, and (b) transmission and reflection spectra of a gold grating for $P=75$ nm, $w=20$ nm, $h=198$ nm, $n_s=1$, and $\theta=0^o$.}
\label{fig2_chap4}
\end{figure}

\begin{figure}
\centering
\includegraphics[width=7.5cm]{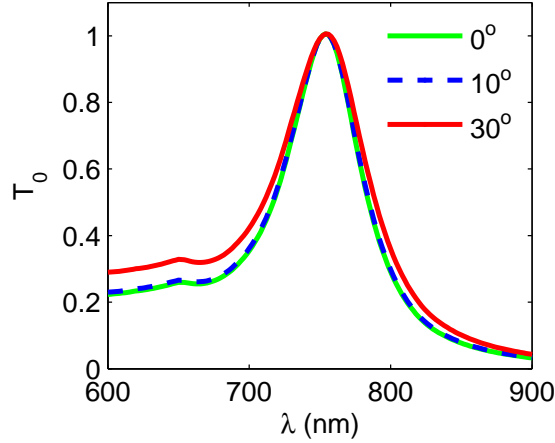}
\caption{Zeroth order transmission at different incidence angle for the same grating parameters of Fig. \ref{fig2_chap4}.}
\label{fig3_chap4}
\end{figure}

To justify the validity of the transmission and reflection models developed in the last section, geometrical grating parameters from Ref. \cite{Rahman2012} are considered. Specifically, $P=75$ nm, $w=20$ nm, $h=198$ nm, $n_s=1$, $\theta=0^o$ and the grating is made of gold are assumed. These parameters are supposed to produce a Fabry-Perot resonance peak at $750$ nm and no transmission peak due to the Wood-Rayleigh Anomaly (WRA) or the surface plasmon resonance is expected in the visible band. Fig. \ref{fig2_chap4}(a) shows the zeroth order transmission spectrum corresponding to these grating geometrical parameters. In this figure, in addition to the data available from Eq. (\ref{eq15_chap4}), results obtained from the exact modal analysis \cite{Sheng:82} with thirteen modes and the analytical model developed by Garcia-Vidal et al. \cite{GarciaVidal2002} have also been included. One can see that, except the amplitude mismatch, the analytical model developed here (Eq (\ref{eq15_chap4})) fits very well with the EMA results. In contrast, as mentioned earlier, the semi-analytical model developed by Garcia-Vidal and co-workers \cite{GarciaVidal2002} completely fails to predict the resonance. However, if one closely looks at the denominator of Eq (\ref{eq15_chap4}), then it is evident that when the FB resonance condition is met ($n_{Eff}^{'}k_0h=j\pi$ with $j=1, 2, 3,..$ \cite{Astilean2000}), $\cos{(n_{Eff}^{'}k_0 h)}=1$ and $\sin{(n_{Eff}^{'}k_0 h)}= 0$. The consequence of the FP resonance on the denominator is that it becomes smaller compared to non-resonance scenarios. This ensures that $T_0$ is maximized and a transmission peak appears. Fig. \ref{fig2_chap4}(b) confirms that at the transmission peak, reflection is reduced as expected. Fig. \ref{fig3_chap4} shows the dependence of the zeroth order transmission on the angle of incidence. It is known that when $P << \lambda$ (in our case $P=75$ nm and $\lambda=750$ nm), the dependence of the FP resonance on $\theta$ is minimal \cite{Porto1999,Cao2002,Tyan:97}. This is confirmed from the results of Fig. \ref{fig3_chap4}. It can also be understood from Eq. (\ref{eq15_chap4}). Elaborately, irrespective of the incidence angle, the absolute value of the numerator of Eq. (\ref{eq15_chap4}) is always $2$. Additionally, when the FP resonance condition is satisfied and $n_s=1$, the $1^{st}$ and the $2^{nd}$ terms within the square bracket of the denominator of Eq. (\ref{eq15_chap4}) become independent of $\theta$ and zero, respectively. Upon these circumstances, Eq. (\ref{eq15_chap4}) simplifies to Eq. (\ref{eq15a_chap4}). From this last equation it can be seen that the incidence angle has no impact on the FP resonance. Further, from Eq. (\ref{eq15_chap4}) it can be understood that a peak related to the FP resonance in the extraordinary optical transmission through a $1$D metallo-dielectric grating is independent of the grating period ($P$). This has been experimentally verified by Pang and coworkers \cite{Pang2007}. Also, it is evident from Eq. (\ref{eq15a_chap4}) that as the grating thickness ($h$) increases (for further discussion on the impacts of $h$ on transmission, see below), the intensity of a peak related to the FP resonance decreases. At this point it is relevant to mention that these are the kind of insights that our model can provide while other existing models of $1$D metallo-dielectric gratings fail to deliver.

\begin{eqnarray}
\small
|T_0|=\frac{1}{\cosh{(n_{Eff}^{''}k_0h)}}
\label{eq15a_chap4}
\end{eqnarray}

The role of the slit width on the transmission intensity can be appreciated through the influence of $n_{Eff}^{''}$ on Eq. (\ref{eq15_chap4}). It is known that $n_{Eff}^{''}$ varies inversely with the slit width ($w$, see Eq. (\ref{eq3_chap4})) or $n_{Eff}^{''}$ increases as $w$ decreases and vice versa. The result of the inverse relationship between $w$ and $n_{Eff}^{''}$ is that as the slit width shrinks $\cosh{(n_{Eff}^{''}k_0h)}$ in Eq. (\ref{eq15_chap4}) increases and hence the transmission intensity diminishes as intuitively expected.

To understand the effects of the grating thickness ($h$) on the overall transmission intensity related to the Fabry-Perot resonance, let us go back to Eq. (\ref{eq13_chap4}) with $\theta=0^o$, $n_s=1$ and $m=0$. Neglecting terms with $\frac{n_{Eff}}{\epsilon_r}$ in the denominator, one gets -

\begin{eqnarray}
\small
\nonumber
T_0&=&\frac{2\exp{(-ik_0h)}}
{2\cos{(n_{Eff}k_0h)}-i\frac{\epsilon_r}{n_{Eff}}\sin{(n_{Eff}k_0h)}}\\
\nonumber
T_0&=&2\exp{(-ik_0h)}/\\\nonumber
&&\textbf{[}(2\cos{(n_{Eff}^{'}k_0h)}-i\frac{\epsilon_r}{n_{Eff}}\sin{(n_{Eff}^{'}k_0h)})\cosh{(n_{Eff}^{''}k_0h)}-\\&&(2\sin{(n_{Eff}^{'}k_0h)}+i\frac{\epsilon_r}{n_{Eff}}\cos{(n_{Eff}^{'}k_0h)})\sinh{(n_{Eff}^{''}k_0h)}\textbf{]}
\label{eq17_chap4}
\end{eqnarray}

It is clear that as $h$ increases both $\cosh{(n_{Eff}^{''}k_0h)}$ and $\sinh{(n_{Eff}^{''}k_0h)}$ in the denominator of Eq. (\ref{eq17_chap4}) rise rapidly and as a consequence $T_0$ is expected to reduce significantly. In other words, if one uses a thicker grating to attain a FP resonance at a desired $\lambda$ then the associated transmission peak will be weaker compared to the grating which has a smaller $h$. This has been experimentally confirmed by other research groups \cite{Ebbesen1998,Pang2007}. Lastly, to understand the role of the grating material (specifically the ridge material) on the transmission through a $1$D grating, inspecting the denominator of Eq. (\ref{eq17_chap4}) can again be useful. Elaborately, for a fixed $w$ and $P$, as $\epsilon_r$ increases, the denominator of Eq. (\ref{eq17_chap4}) increases and consequently $T_0$ is supposed to decrease proportionately. Alternatively, it can be concluded that if one considers two metals of different dielectric constants to get a FP resonance with the minimum possible grating thicknesses at a desired $\lambda$ (a metal with a higher $n_r$ requires a thicker grating to fulfil the FP condition, see equations (\ref{eq3_chap4})) then the metal which has a higher $|\epsilon_r|$ (in particular a metal which has the higher imaginary dielectric constant) will produce a weaker transmission peak compared to the metal which has a smaller $|n_r|$. Again this has been verified \cite{Garcia2007}.

\begin{figure}
\centering
\includegraphics[width=7.5cm]{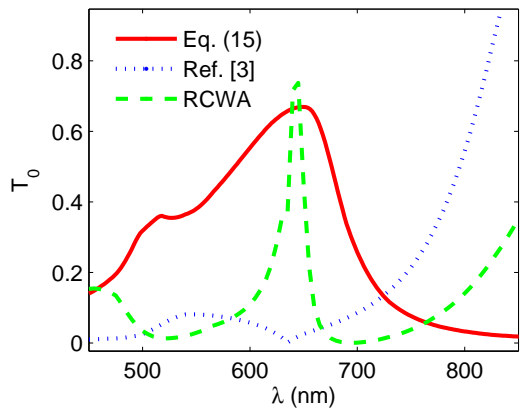}
\caption{Zeroth order transmission for $P=418$ nm, $w=50$ nm, $h=209$ nm, $n_s=1.52$, $\theta=0^o$ and the ridge metal is gold.}
\label{fig5_chap4}
\end{figure}

Before concluding, one more example is considered. In this example, we consider $P=418$ nm , $w=50$ nm, $h=209$ nm, and a gold grating \cite{Palik85} on a glass substrate ($n_s=1.52$). These grating geometrical parameters are supposed to produce a transmission peak at around $635$ nm. Equivalently, the grating should work as a color filter by transmitting the red light while blocking all other colors \cite{Rahman2012A}. Corresponding to these parameters, a Wood-Rayleigh anomaly and a FP resonance should occur at $635$ nm. Fig. \ref{fig5_chap4} shows the zeroth order transmission spectra obtained using the model ( Eq. (\ref{eq15_chap4})) developed in this article, the model of Lalanne et al. \cite{Lalanne2000A}  and the Rigorous Coupled Wave Analysis (RCWA). It can be observed that, like the RCWA, the simple analytical model developed in this article, despite some amplitude mismatch, can predict the wavelength and the profile of the resonance while the model of Ref. \cite{Lalanne2000A} completely fails. According to us the main reason behind the failure of Lalanne and co-workers' model of $1$D grating is the PEC assumption that they have made in deriving their model. As mentioned earlier, the PEC assumption is valid deep inside the infrared and the microwave regions of the electromagnetic spectrum and that is where this model matches very well with the RCWA results (such as those of Ref. \cite{Cao2002}). In the visible band of the electromagnetic spectrum, the perfect electrical conductor assumption is invalid for metals like gold, silver and copper. Consequently, results produced using the model of Lalanne et al. are inaccurate. Lastly, it is noticeable from Fig. \ref{fig5_chap4} that the line-width of the peak of the analytical model developed here is broader than that obtained from the RCWA. Such a broader peak is normally a characteristic of the FP resonance. Since the model developed here is insensitive to the Wood-Rayleigh anomaly, the effects of the WRA is not reflected in the spectrum obtained using Eq. (\ref{eq15_chap4}). Consequently, a mismatch between the result generated using the RCWA, that includes the influences of both WRA and FP mechanisms, and the data available from Eq. (\ref{eq15_chap4}) is expected.

\section{Summary}

In conclusions, an analytical model of light transmission and reflection through $1$D metallo-dielectric gratings based upon the exact modal analysis has been developed. This model can accurately predict a transmission peak due to the Fabry-Perot resonance. It has also been theoretically proved that transmission intensity of a peak related to the Fabry-Perot resonance decreases with the increasing grating thickness and the refractive index of the ridge material. Further, when the grating period is much smaller compared to the incident wavelength, the model presented in this article can capture the variation of transmission intensities as the incident angle changes. Limitation of this model is that it can not stipulate a transmission peak due to the Wood-Rayleigh anomaly.


\bibliographystyle{elsarticle-num}



\end{document}